# Collective Properties of Cellular Identity: a computational approach

Bradly Alicea[1,2]
bradly.alicea@ieee.com
[1] Cellular Reprogramming Laboratory, Michigan State University  [2] Orthogonal Research

**ABSTRACT**

Cell type (e.g. pluripotent cell, fibroblast) is the end result of many complex processes that unfold due to evolutionary, developmental, and transformational stimuli. A cell's phenotype and the discrete, *a priori* states that define various cell subtypes (e.g. skin fibroblast, embryonic stem cell) are ultimately part of a continuum that may predict changes and systematic variation in cell subtypes. These features can be both observable in existing cellular states and hypothetical (e.g. unobserved). In this paper, a series of approaches will be used to approximate the continuous diversity of gene expression across a series of pluripotent, totipotent, and fibroblast cellular subtypes. We will use a series of previously-collected datasets and analyze them using three complementary approaches: the computation of distances based on the subsampling of diversity, assessing the separability of individual genes for a specific cell line both within and between cell types, and a hierarchical soft classification technique that will assign a membership value for specific genes in specific cell types given a number of different criteria. These approaches will allow us to assess the observed gene-expression diversity in these datasets, as well as assess how well *a priori* cell types characterize their constituent populations. In conclusion, the application of these findings to a broader biological context will be discussed.

## INTRODUCTION

What defines a phenotype? While this question is central to evolutionary biology, it is less central to the study of cell fate and cellular reprogramming. In theoretical evolutionary biology, the question has been understood as a mapping of genotype to the phenotype (Stadler, 2006; Wagner, 2005). While this is difficult to do in terms of experimental biology (Rockman, 2008), there are ways to approximate this relationship using computational tools. These analyses generally focus on normative measurements of gene expression and subsequent inferences of function. This can provide insights into the properties of and mechanisms behind physiological changes themselves. However, this approach is far less satisfying for questions revolving around cellular identity, which is focused more on underlying patterns and tendencies in the data.

In the context of cellular transformation and reprogramming, phenotype is the end result of many complex processes. This is both influenced by and produces a diversity of outcomes, only some of which are essential to defining the cellular state. Our goal is to extract features from our datasets representative of pluripotent, totipotent, and fibroblast diversity that are either observable among existing cellular states or hypothetical (e.g. unobserved). In cases where these cellular states are truly discrete, a Hidden Markov Model (HMM) can be used (Eddy, 2004). However, our interest is in the continuous nature of cellular diversity. This is an important aspect of transforming cells, as the unfolding of transformative processes such as oncogenesis (Luo and Elledge,2008) or cellular reprogramming (Soufi, Donahue, and Zaret, 2012) are influenced greatly by this variation.

We have chosen a series of fibroblasts and pluripotent cell lines to understand what defines a cell type. This has been studied both philosophically and empirically in pluripotent cell types using the concept of stemness. Typically, cell types are given an *a priori* or top-down identity through consensus and observation: a collection of attributes such as phenotypic markers, gene expression, and microenvironmental niche usually go into defining cells of various origins in the same species as a fibroblast or a stem cell (Hamilton, Pantelic, Hanson, and Teasdale, 2007; Sung, Park, and Kim, 2011; Ramunas et.al, 2007).

We will then uncover strategies for revealing the fundamental structure and transitional features of cellular transformation using a database of exemplar cell populations and a number of candidate computationally-oriented strategies. To accomplish this, we will rely on three interrelated tests: the computation of distances based on the subsampling of diversity, assessing the separability of individual genes for a specific cell line both within and between cell types, and a hierarchical soft classification technique (Babuska, 1998) that will assign a membership value for specific genes in specific cell types given a number of different criteria (Jin and Wang, 2009).

A series of analyses will be done a number on previously-collected datasets (see Methods) to ask two related questions. The first question is: given high-throughput gene expression data, what is the observed diversity within and between cell types? The second question is: how well do *a priori* cell types and cell line definitions characterize their constituent populations? Finally, to put these results into a context of transitions from one cellular phenotype to another, we can use an time-series representing embryonic gene expression to understand changes in gene expression that occur during active changes in cellular phenotype and pluripotency.

**Analysis overview**

The first approach involved a model of objective distances[1] both within and between fibroblasts and pluripotent cell lines. This was done on the fold-change transformed data using two criteria: absolute pairwise distance between cell lines greater than 2.0, and an absolute pairwise distance between cell lines less than 0.05. These two criteria define the extremes and the invariant regions of the fold-change distribution, respectively, for every cell line (demonstrated in Figure 1 using an idealized Gaussian). The second approach extended this basic model to a more specific querying technique based on the intersection of percentile and cell type for a given cell line. This can be used to generate profiles which characterize diversity at very specific points in the cell type space.

Due to the exploratory nature of the first two techniques, two additional approaches were developed and brought to bear on this problem. The first of these is a test of separability between each gene for all cell line pairings. This provides as indication of the structure inherent for a certain gene or set of genes across cell type diversity. The final approach involves using a hierarchical fuzzy classification model to better understand the internal structure of non-seperable relationships within each cell type. An overview of how the analyses will proceed is shown in Figures 1 and 2.

---

[1] the probe values for multiple microarrays representing a single cell line were averaged together. The entire range of probe values for each cell line is then transformed into a set of z-scores.

*Separability.* The concept of separability (Hanan, 2006) is a statistical phenomenon that describes the independence of two or more distributions. Separability is an underlying concept of data decomposition, and occurs when subsets of data can be distinguished by a kernel. Determining seperability can also act as a decision rule for confirming the *a priori* boundaries of verbal (or semantic) classifications. Further information can be found in the Methods section.

*Soft classification.* The concept of soft (or fuzzy) classification (Chen and Pham, 2001) is a statistical method that allows us to assess the membership of an element (in this case a gene) in a set. Soft classification provides a degree of membership, which is similar to a probability, except that any one element can belong to two or more sets simultaneously. This can provide a basis for classification using multiple criteria. Further information can be found in the Methods section.

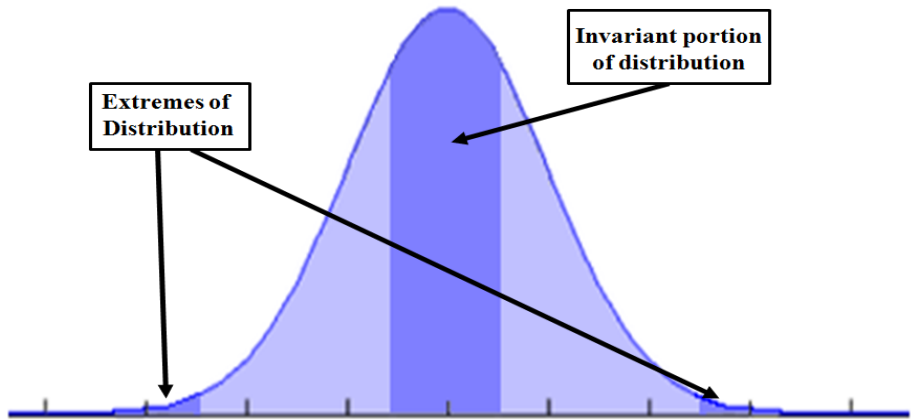

**Figure 1. Cartoon describing the most invariant and most extreme criteria.**

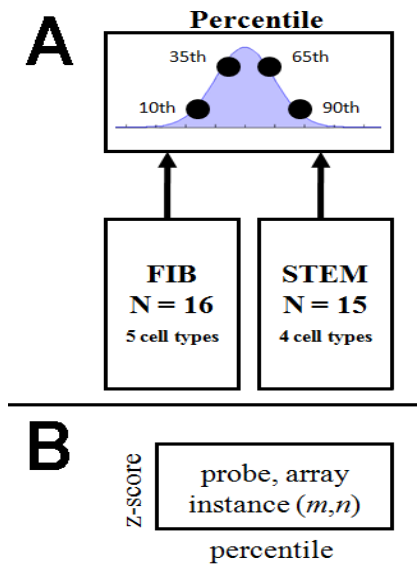

**Figure 2. Cartoon describing the workflow for mapping variation in both the fibroblast and pluripotentcy datasets.**

# RESULTS

**Objective distance criteria**

     The first test involves calculating distances between different cell lines and cell types. Distances used in this test were determined using two criteria. The first defines the most standard or invariant genes across either pluripotent or fibroblast cell types. The second defines the most extreme genes, or the genes that change the most across either pluripotent or fibroblast cell types. Each test provides a hierarchical model for which we can test the idea that fibroblast and pluripotent cell types are indeed separate entities. Therefore, distances have been calculated for cell lines between types. This can be summarized using diagrams shown in Figures 3 and 4.

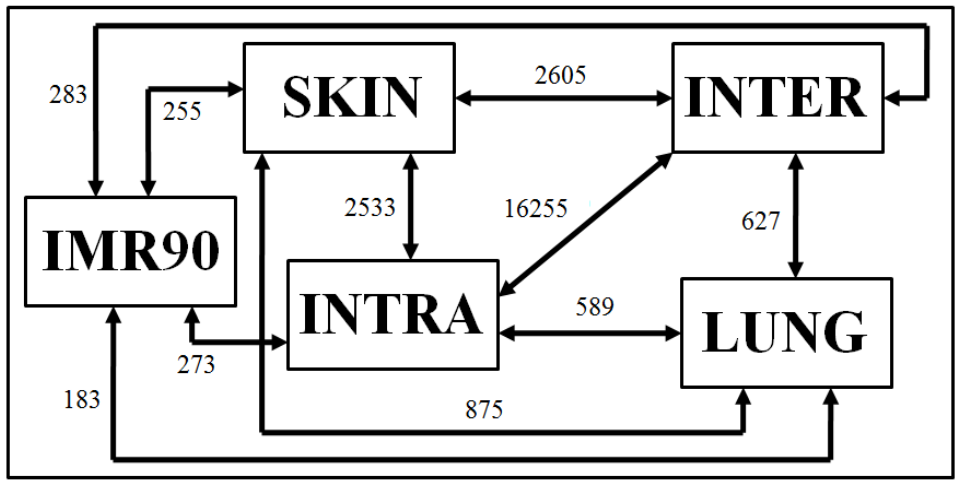

**Figure 3. Map of distances between fibroblast cell lines, within cell type. Whole numbers on arc denote number of genes within the most standard/invariant criterion for a given cell line pair. Total of 12,362 probes included for analysis.**

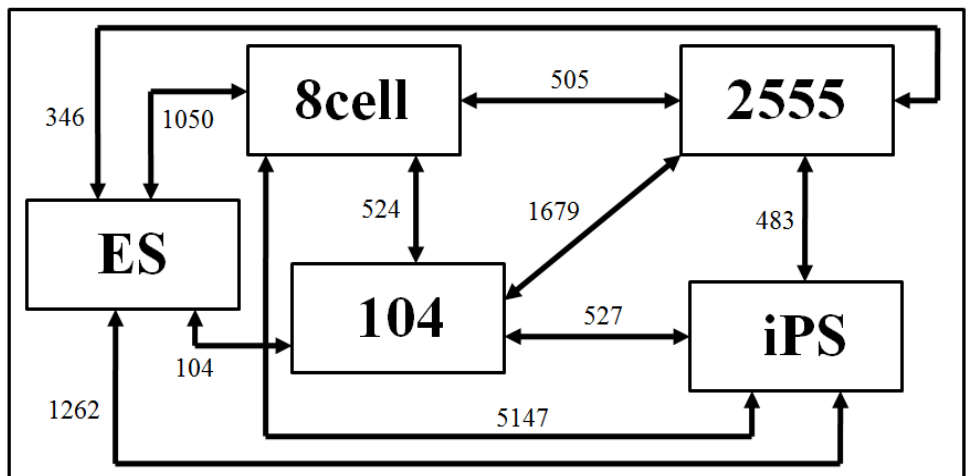

**Figure 4. Map of distances between pluripotent cell lines, within cell type. Whole numbers on arc denote number of genes within the most standard/invariant criterion for a given cell line pair. Total of 12,362 probes included for analysis.**

Comparisons between different types of breast-derived fibroblast in Figure 3 reveal the highest number of standard/invariant genes. Another trend apparent in Figure 4 is the transitive relationship between iPS, 8 cell, and ES cell lines. While the relationship between our ES-iPS (1262) and ES-8 cell (1050) yield a similar number of genes, the iPS-8 cell relationship (5147) yields a number which is at least 4-fold higher. This suggests that iPS cells and 8-cell stage of development can yield many similarities. On the other hand, the other iPS cell lines in our analysis (104, 2555) yield about half the number of similar genes than do the ES-iPS and ES-8 cell relationships. This may suggest that iPS-style pluripotency, and indeed pluripotency in general, is achieved and regulated using a mosaic of different genes, the identity of which being context-dependent.

**Table 1. Number of genes with the most extreme criterion for a given cell line pair. Criterion: absolute difference > 2.0 when comparing fibroblast (columns) and pluripotent cells (rows) by their *a priori* states. Total of 12,362 probes included for analysis.**

|  | iPS (GEO) | iPS (CRL) | ES (GEO) | 8-cell (GEO) |
|---|---|---|---|---|
| SKIN (GEO) | 46 | 376 | 210 | 350 |
| INTRALOBULAR (GEO) | 1658 | 1290 | 1840 | 1665 |
| INTERLOBULAR (GEO) | 1650 | 1300 | 1843 | 1686 |
| WI-38 (GEO) | 29 | 320 | 182 | 298 |
| IMR-90 (CRL) | 620 | 622 | 811 | 818 |

**Table 2. Number of genes with the most standard/invariant criterion for a given cell line pair. Criterion: absolute difference < 0.05 when comparing fibroblast (columns) and pluripotent cells (rows) by their *a priori* states. Total of 12,362 probes included for analysis.**

|  | iPS (GEO) | iPS (CRL) | ES (GEO) | 8-cell (GEO) |
|---|---|---|---|---|
| SKIN (GEO) | 1455 | 521 | 977 | 597 |
| INTRALOBULAR (GEO) | 399 | 386 | 402 | 396 |
| INTERLOBULAR (GEO) | 394 | 438 | 398 | 373 |
| WI38 (GEO) | 1717 | 470 | 1437 | 720 |
| IMR-90 (CRL) | 465 | 4120 | 402 | 552 |

When considering the relationships between cell types and using a criterion of > 2.0 (Table 1), we can focus on the very small and very large relationships relative to the average numbers of genes for the matrix. Notably, the pairwise relationships involving WI-38-iPS and skin-iPS cells share a common signature for less than 10% of all genes sampled. Perhaps there are some mechanisms which are upregulated to determine fibroblast identity are not upregulated in skin cells as they are in other fibroblast types.

The same thing can be done for a criterion of < 0.05 (Table 2). In this matrix, we can see that IMR-90 and the iPS lines from the CRL Lab share 33% of their genes. It is noteworthy that these two lines represent a transformation of the same cell population. In addition, the skin-iPS, WI-38-iPS, and WI-38-ES pairwise relationships all reveal a common expression signature for more than 10% of genes sampled. Perhaps WI-38, which is derived from embryonic liver, shares some developmental-oriented mechanisms with both iPS and ES cells.

The distances within and between cell types, even using multiple criteria, do not tell the whole story. The distances presented in Figures 2 and 3 and Tables 1 and 2 are based on cumulative (e.g. mean, sum) values for each cell line, without regard for variation within a cell line. To address this, we estimated distributions for each cell line based on all available replicates. The ranges of these distributions are then compared with each other to determine their absolute and relative separability.

**Separability tests**

Separability tests were done using two criteria (see Methods). As a statistical property, separability defines the independence of any two distributions. Therefore, if a given gene for any two cell lines is separable, the activity of that gene should not be correlated and thus contributes to observed phenotypic differences. A separate test for the effects of outliers on the result (e.g. false positives) revealed that such effects were absent. Tests were conducted at two different levels of analysis: among cell lines between different cell types, and across cell lines with a cell type.

The first test involves examining absolute separability for pairs of cell lines within and between cell types. This involves a discrete test of each gene in a pairwise fashion across the range of diversity. Results among lines but between cell types are shown in Figure 5, while results across the Embryo time-series are shown in Supplementary Figure 1 (S1).

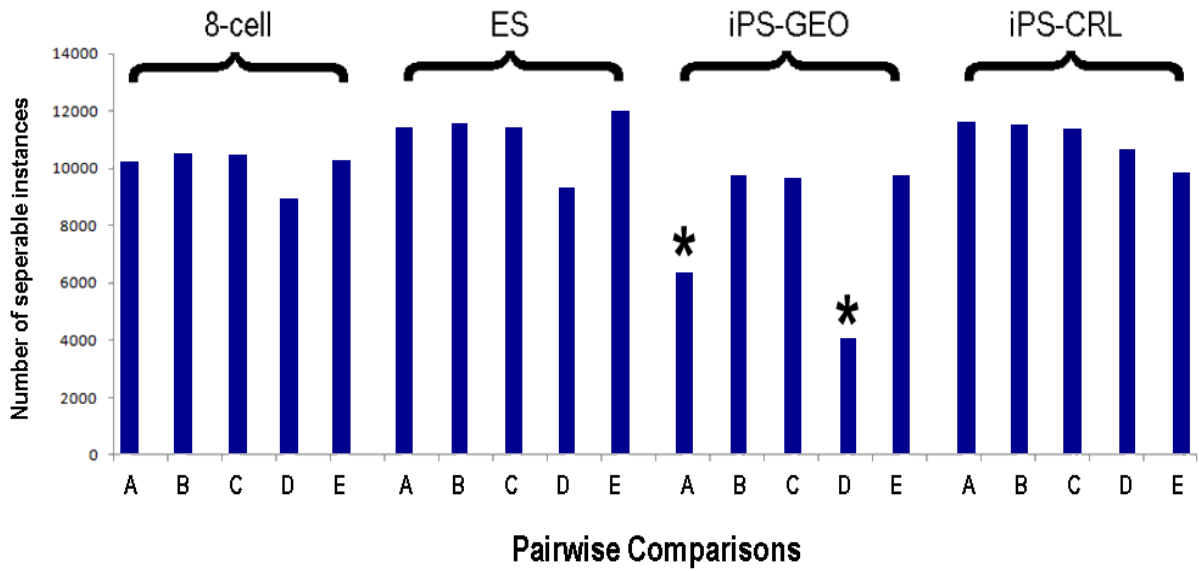

**Figure 5. Graph showing number of separable genes for pairwise comparisons across cell types. Total of 12,362 probes included for analysis. From left (each comparison): A) Skin, B) Intra-breast, C) Inter-breast, D) WI-38, E) IMR-90.**

As is shown in Figure 5, there is uniformity in the number of seperable instances when comparing fibroblast and pluripotent cell lines except for a few cases. The two cases of most interest are the iPS-GEO comparison with Skin fibroblasts and WI-38 (lung) cells.

The second test involves examining relative separability for all cell lines within a given cell type. This involves a continuous test of each gene's separability for all possible pairings within a given cell type. This comparison is cell type independent, and so provides a signal that can be compared between cell types (shown in Figure 6).

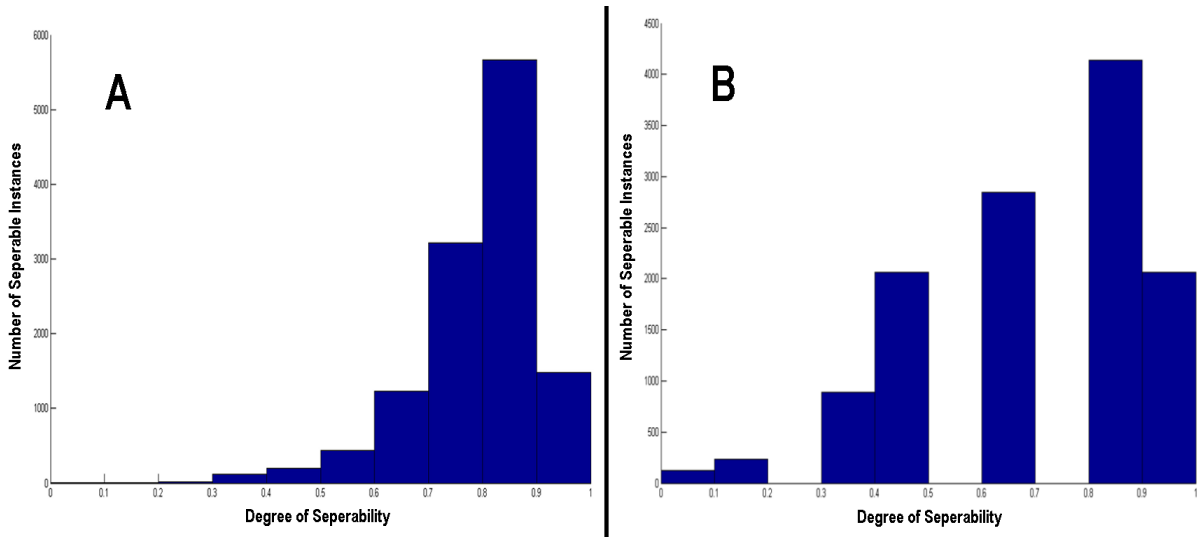

**Figure 6. Graph showing histogram of relative separability (frequency of genes for each number of seperable instances).**

**3rd-order fuzzy classification model**

Soft classification using fuzzy criteria allows us to understand more deeply the relative components of cell identity as characterized by gene expression. Fuzzy sets (kernels) have membership functions. A given site (gene, cell type combination) can have simultaneous memberships in multiple kernels (functions with a specific membership criterion). This will allow us to further understand exactly how pluripotent and fibroblast cell types are differentiable by marker, and by extension how their state is regulated (e.g. through mechanisms common to each cell type, or through a mosaic mechanism that recruits genes independently).

The design of a 3rd order fuzzy classification model is shown in Supplementary Figure 2 (S2). Four kernels were computed for all data (hi, low, Q, and dual – see Methods for definitions). Three tests were conducted to examine the relationship between genes 1) within a single cell line but across microarray instances, 2) for a gene in a single cell line but across a given cell type, and 3) within a single cell line but across another cell type.

A principle components analysis (PCA) was conducted on all kernels, but yielded no interesting results. A more appropriate approach was then tried. The results shown in Figures 7, 8 and Supplemental Figures 3-5 (S3-S5) show a non-decompositional frequency analysis based on the "dual" criterion. The resulting plot is referred to as a membership density function, which compares the memberships over a range of values.

A 3rd-order fuzzy classification of embryo time-series data was also conducted used to better understand how the nature of separability between dependent datapoints (e.g. the activity of a specific gene across stages of embryonic development). Figures 7 and 8 demonstrate the results of these analyses.

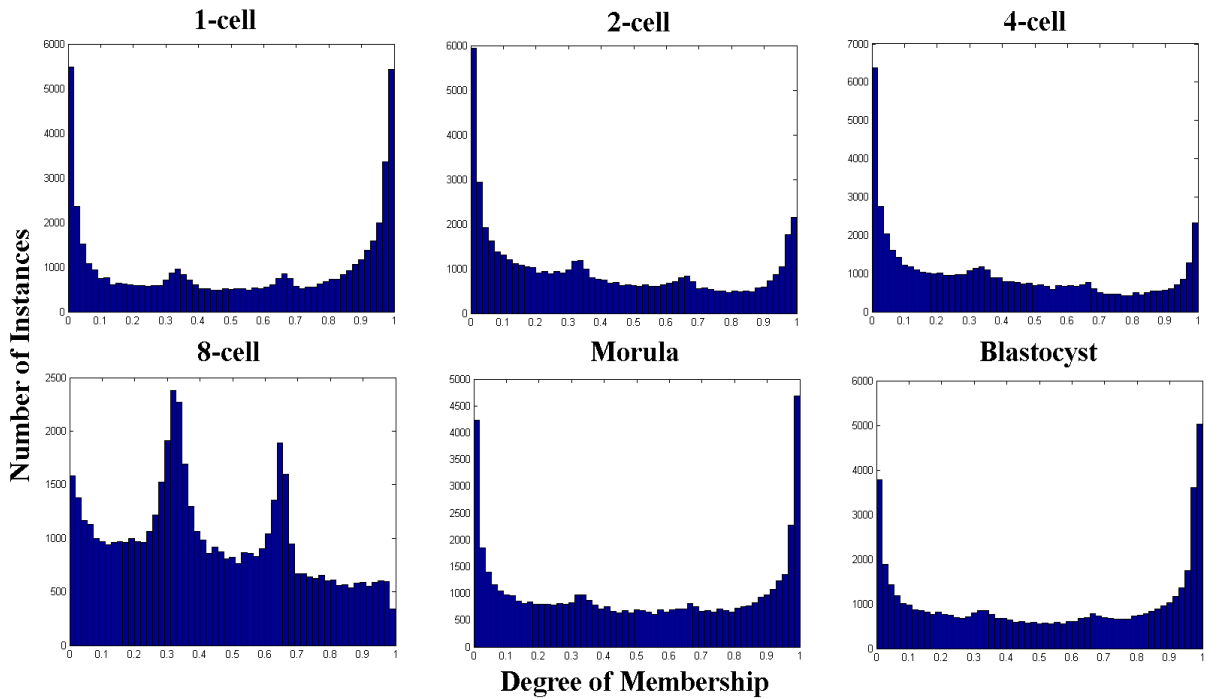

**Figure 7. The membership density functions for each cell line in embryo time-series dataset (1 cell through Blastocyst). Histogram of 50 bins, total of 54,675 --- probes used in the analysis.**

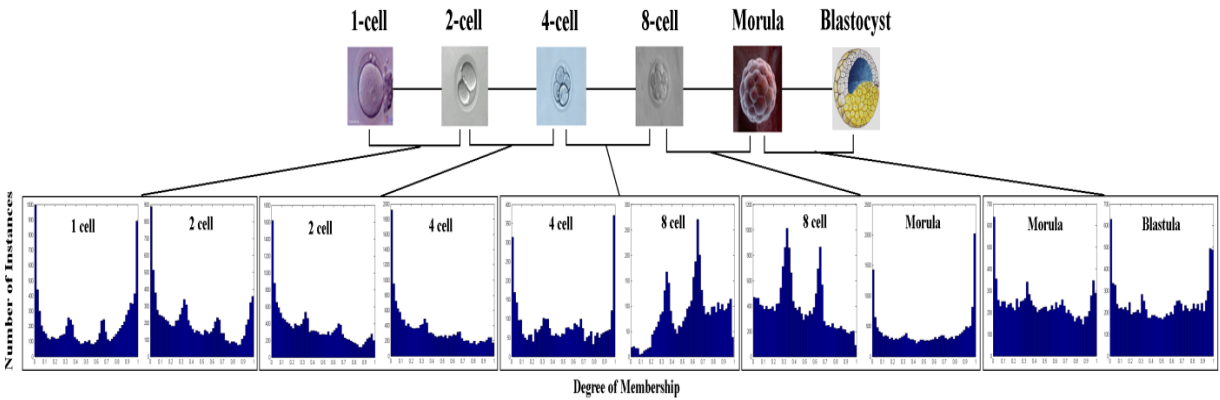

**Figure 8. Graph showing the membership density functions of all seperable sites based on pairwise comparisons of cell lines. Histogram of 50 bins, total of 54,675 --- probes used in the analysis.**

For the embryo time series, we can see a number of interesting trends. Figure 7 shows the membership density function for all genes in the analysis. In this analysis, we can see what looks like an inverse bell curve with two internal peaks at 0.35 and 0.70. The "tails" (categories closest to 0.0 and 1.0) fluctuate across the time series.

When examining the membership density functions (Figure 7) for each cell line, two interesting features stand out. One is the downward shift in the tail near the 1.0 membership for the 2-cell and 4-cell state, along a similar downward shift in both tails (for 0.0 and 1.0 values) for the 8-cell state. Secondly, the 8-cell state exhibits a corresponding upward shift in the internal peaks for the 8-cell state.

Finally, we asked whether a similar outcome would result if only known seperable probes were used. The membership density function (dual kernels from the soft classification) data were resampled with respect to whether or not the z-score transformed expression values were previously classified as seperable (see Figure 8). For each pairing adjacent in time (e.g. 1-cell vs. 2-cell), two graph were produced, one for each cell type in the pair. In this way, 2-cell, 4-cell, 8-cell, and Morula data were resampled twice. A similar pattern of membership for each cell type is revealed, complete with accentuated internal peaks for the 8-cell samples. In the case of 8-cell and Morula data, the double resampling shows large but consistent differences between them (Figure 8).

## CONCLUSIONS

Now we will review the results for the seperability and soft classification tests, and place them in context. The first set of results (seperability test) will define the range of each cell line and set up the soft classification analysis. In conclusion, these results will be framed in the context of cellular reprogramming, and how they may more generally elucidate the process of phenotypic transformation.

**Separability tests**
The results in Figure 5 show that independent pluripotent cell lines are largely (70-80%) seperable of independent fibroblast cell lines using an absolute test of separability. There are a couple of exceptions to this, none of which can be easily explained. The results in Figure 6 show that cell lines with a pluripotent *a priori* identity are separable in a way that is much more variable than among fibroblast cell lines. This suggests that pluripotent cell lines maintain their cell type using myriad sets of genes and molecular mechanisms, which does not seem to hold true for the examined fibroblast cell lines. This result can be compared with separability between dependent cell lines. In this case, the degree of separability is far lower (from 10-35%). However, it appears that aggregate patterns revealed through soft classification

**Soft (fuzzy) classification**
While the soft classification technique was envisioned as a way to clarify what is revealed by the separability metric, the results are decidedly mixed. One outcome is that we can observe shifts in the distribution when a specific cell type is embedded in a broader background. This is particularly true for the dual criterion. However, the true meaning of these patterns is far from certain. The outcome of revealing broad patterns is similar to an approach called genomic signal processing Shmulevich and Dougherty, 2007), which can be used to explore the differences between different cell types (Alicea, 2013).

**Applications to the study of phenotypic transformation**
There are a number of outcomes with application to cellular transformation observed in disease, evolution, and bioengineering. The main relevant finding involves the role of individuality in classifying cell as a "type". This is one reason why we did not use more conventional machine learning approaches to classifying the data. Making the connection between the available data and what is going on inside the cell requires a better understanding of the regulatory mechanisms behind cell type.

While it is beyond the immediate scope of this paper, the nature of cellular identity, memory, and transcriptional regulation can help frame the results of this paper in context. Specifically, this may help us understand the more subtle patterns found in gene expression variation within and between cell types. In a review focused on the immune system, Fisher (2002) presents the idea that cellular responses to environmental signals involve a "working memory" buffer that keep cellular state stable. Bentolia (2005) argues more generally for a "live memory" mechanism that does not directly involve genetic information, but does provide a selective mechanism that imposes stability on otherwise volatile biological processes. In both cases, memory is selective and retains information about the current cellular state.

Ivanova et.al (2002) provides a definition of common molecular signatures that define a cellular state. In the case of stem cells, we should expect to find an overlapping set of gene products that is shared between species and cellular fates (e.g. embryonic and neural stem cells). Does this serve a memory mechanism? Enver et.al (1998) suggests that cellular memory is based at least in part on context-dependent mechanisms. Even though the same pathway can be activated in two different cells, pathway activation can lead to multiple outcomes. This may be due to subtle patterns of activation resulting from higher-order regulatory mechanisms.

In yeast, there are many known positive and negative regulatory elements that require a series of binary decisions to be made (Acar, Bacskei, and van Oudenaarden, 2005). In some cases, this may result in stable expression levels, while in others it can result in enhanced cellular memory. More importantly, these decisions result in a complex set of regulatory loops. In cases where cells transition from one type (e.g. state) to another, there are two predominant types of regulatory mechanism: feedforward cascades that cooperatively execute a transformative process, and mutually-exclusive cell-specific decisions. This mutual exclusivity can be better understood by extending our seperability and soft classification analyses to other contexts, particularly samples of gene expression from across the process of phenotypic specification.

## METHODS

**Separability metric**

For any two cell lines, A and B, a gene *n* is seperable ($S_n = 1$) when either of two conditions can be met

$$\mathbf{S_n} = \begin{cases} |\min(A) - \min(B)| < |\min(A) - \max(A)|, 1 \\ |\min(B) - \max(A)| < |\min(B) - \max(B)|, 1 \\ \text{else}, 0 \end{cases} \qquad [1]$$

To test for false positives, the z-scores for each line were calculated from the normalized microarray data. All replicates of each gene, cell line pair was considered an outlier if any values were greater than 2.0 or less than -2.0. Associated code can be accessed on Github: https://github.com/balicea/collective-properties-cellular-identity

**Soft (fuzzy) classification**

There are four soft categories used to classify fibroblast, pluripotent, and embryonic gene expression: the *Hi* criterion, the *Lo* criterion, the *Q* criterion, and the *Dual* criterion. These kernels define different aspects of this diversity, and provide several potential signals for analysis.

*Hi criterion.* The high criterion classifies the degree of membership of a given gene (instance) within a range of values. This degree of membership ranges from 0.0 to 1.0, with the minimum value in this range being 0.0, and the maximum value in this range being 1.0. This can be defined mathematically as

$$\text{HI} = \frac{A_i}{\text{MAX}(A) - \text{MIN}(A)} \qquad [2]$$

*LO criterion.* The low criterion classifies the degree of membership of a given gene (instance) within a range of values. This degree of membership ranges from 0.0 to 1.0, with the maximum value in this range being 0.0, and the minimum value in this range being 1.0. This can be defined mathematically as

$$\text{LO} = \frac{1 - A_i}{\text{MAX}(A) - \text{MIN}(A)} \qquad [3]$$

*Q criterion.* The Q criterion classifies the degree of membership of a given gene (instance) within a range of values. This degree of membership ranges from 0.0 to 1.0, with both the minimum *and* maximum value in this range being 0.0, with a range of values in the middle of the distribution (e.g. support region) being assigned a value of 1.0. This support region, or flat-topped peak, characterizes all z-scores on the interval 1.0 to -1.0. This can be defined mathematically as

$$Q = \begin{cases} \frac{A_i}{-(\text{MIN}(A) - (-1.0))}, x: \to 1.0 \\ (-1.0) < A_i > (1.0), \ x = 1.0 \\ \frac{1 - A_i}{\text{MAX}(A) - 1.0}, x: \to 1.0 \end{cases} \qquad [4]$$

*Dual criterion.* The dual criterion classifies the degree of membership of a given gene (instance) within a range of values. This degree of membership ranges from 0.0 to 1.0, with both the minimum *and* maximum value in this range being 0.0, with a single values in the middle of the distribution (e.g. support region) being assigned a value of 1.0. This results in a peaked function, distinct from flat-top of the Q criterion (see Supporting Figure 1 -- S1). This can be defined mathematically as

$$\text{DUAL} = \begin{cases} \frac{A_i}{-(\text{MIN}(A) - 0)}, x: \to 0.5 \\ \frac{1 - A_i}{\text{MAX}(A) - 0}, x: \to 1.0 \end{cases} \qquad [5]$$

These four criteria were used to classify data within a cell line, between cell lines, and across cell lines. Due to their complex nature, DUAL kernels are also defined by an additional set of criteria.

$$\text{Poss}(A,X) = \sup_{x \in X} [A(x)\, tX(x)] \quad \quad [6]$$

$$\text{DUAL}(x,y) = |\max(y) - \min(x)| - \left[\frac{1}{\text{median}(x) - \min(y)}\right] \quad \quad [7]$$

Associated code can be accessed on Github: https://github.com/balicea/collective-properties-cellular-identity

**Re-analyzed Datasets**

All analytical work is done using MATLAB, Excel, and R. Thirty-one (31) previously acquired human microarray studies that encompass fibroblast and pluripotent cell type diversity, and embryo time-series diversity. Pre-processing of microarray data is done using the AMP (automated microarray pipeline) tool located at http://compbio.dfci.harvard.edu/amp/. Microarray data are normalized using the RMA (robust multi-chip average) method. *n*-fold expression values are derived by dividing the individual values for each probe into the mean of the dataset. *z*-score transformations are also used to normalize each set of probes, and serves as an alternative to the n-fold expression criterion. Forty-six (46) previously acquired human microarray studies that encompass fibroblast and pluripotent cell type diversity and embryo time-series diversity shown in Table 3.

**Pluripotency datasets.** For ES cells (GEO database accession numbers: GSM194307, GSM194308, GSM194309) lines from Avery et.al (2008) are used. These samples are derived from human tissue, and used the Affymetrix Human Genome U133 Plus 2.0 platform. For one set of iPS cell lines (GEO database accession numbers: GSM245339, GSM245341, GSM245442, GSM257520, GSM257339, GSM257524) lines from Masaki et.al (2007) are used. These samples are derived from human tissue, and used the Affymetrix Human Genome U133 Plus 2.0 platform. For the set of 8 cell embryonic samples (GEO database accession numbers: GSM456652, GSM456653, GSM456654) lines from Xie et.al (2011) are used. These samples are derived from human tissue, and used the Affymetrix Human Genome U133 Plus 2.0 platform. The iPS lines (104, 21010, 2555) each consisted of microarrays collected at the Cellular Reprogramming Laboratory, Michigan State University. These samples are derived from human tissue, and used the Affymetrix Human Genome U133 Plus 2.0 platform.

**Fibroblast datasets.** For Skin fibroblasts (GEO database accession numbers: GSM301264, GSM301265, GSM301266), the control condition from Duarte, Cooke, and Jones (2009) are used. These samples are derived from human tissue, and used the Affymetrix Human Genome U133 Plus 2.0 platform. For mammary gland intralobular fibroblasts (GEO database accession numbers: GSM309434, GSM309438, GSM309450, GSM309452) and interlobular fibroblasts (Accession numbers: GSM309430, GSM309436, GSM309441, GSM309451), the control condition from Fleming et.al (2008) are used. These samples are derived from human tissue, and used the Affymetrix Human Genome U133A platform. For lung fibroblasts (WI-38 cell line, GEO database accession numbers: GSM484752, GSM484813, GSM484814, GSM484815), the DMSO-exposed vehicle control condition from Dreij et.al (2010) was used. These samples are derived from human tissue, and used the Affymetrix Human Genome U133 Plus 2.0 platform. The untreated IMR-90 line consisted of a single microarray collected at the Cellular Reprogramming Laboratory, Michigan State University. These samples are derived from human tissue, and used the Affymetrix Human Genome U133 Plus 2.0 platform.

**Table 3: Source data for the pluripotent and fibroblast models of cellular state based on gene expression**

| Tissue Origin/Species | Number of Microarrays | Model |
|---|---|---|
| Skin, Human | 3 | Fibroblast |
| Intralobular Breast, Human | 4 | Fibroblast |
| Interlobular Breast, Human | 4 | Fibroblast |
| WI-38, Human | 4 | Fibroblast |
| IMR-90, Human | 1 | Fibroblast |
| ES cell culture, Human | 3 | Pluripotent |
| iPS cell culture, Human | 6 | Pluripotent |
| iPS cell culture (CRL lab), Human | 3 | Pluripotent |
| 1 cell embryo, Human | 3 | Pluripotent |
| 2 cell embryo, Human | 3 | Pluripotent |
| 4 cell embryo, Human | 3 | Pluripotent |
| 8 cell embryo, Human | 3 | Pluripotent |
| Morula, Human | 3 | Pluripotent |
| Blastocyst, Human | 3 | Pluripotent |

**Embryonic time-series datasets.** For the 1-cell through the blastocyst time points (6 total), three microarrays per timepoint are used. The following lines from Xie et.al (2011) are used: 1-cell (GEO database accession numbers: GSM456643, GSM456644, GSM456645), 2-cell (GEO database accession numbers: GSM456646, GSM456647, GSM456648), 4-cell (GEO database accession numbers: GSM456649, GSM456650, GSM456651), 8-cell (GEO database accession numbers: GSM456652, GSM456653, GSM456654), Morula (GEO database accession numbers: GSM456655, GSM456656, GSM456657), Blastula (GEO database accession numbers: GSM456658, GSM456659, GSM456660). All samples are derived from human tissue, and used the Affymetrix Human Genome U133 Plus 2.0 platform.

## Ackowledgements

I would like to thank the expertise of the Cellular Reprogramming Laboratory at Michigan State University (particularly Dr. Jose Cibelli) and Dr. Hasan Otu from Bilgi University, Turkey for shaping the content and direction of this paper. I would also like to thank the primary producers of the datasets used in this paper (see References for more information).

**SUPPLEMENTAL FIGURES**

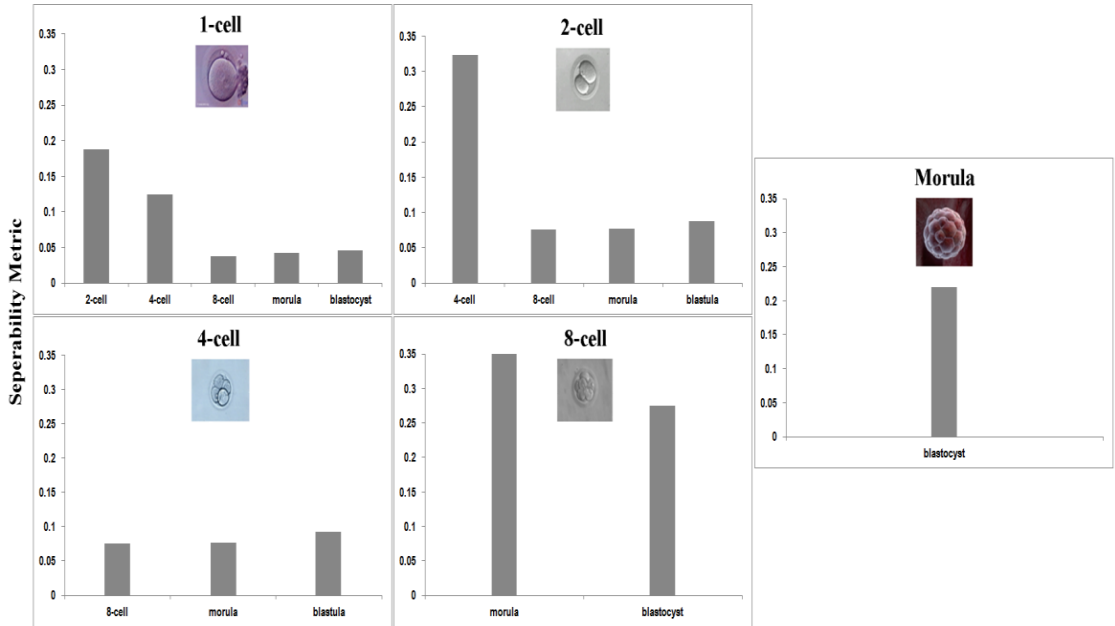

**Figure S1.** Percentage of seperable probes across all possible stages of the embryo time series.

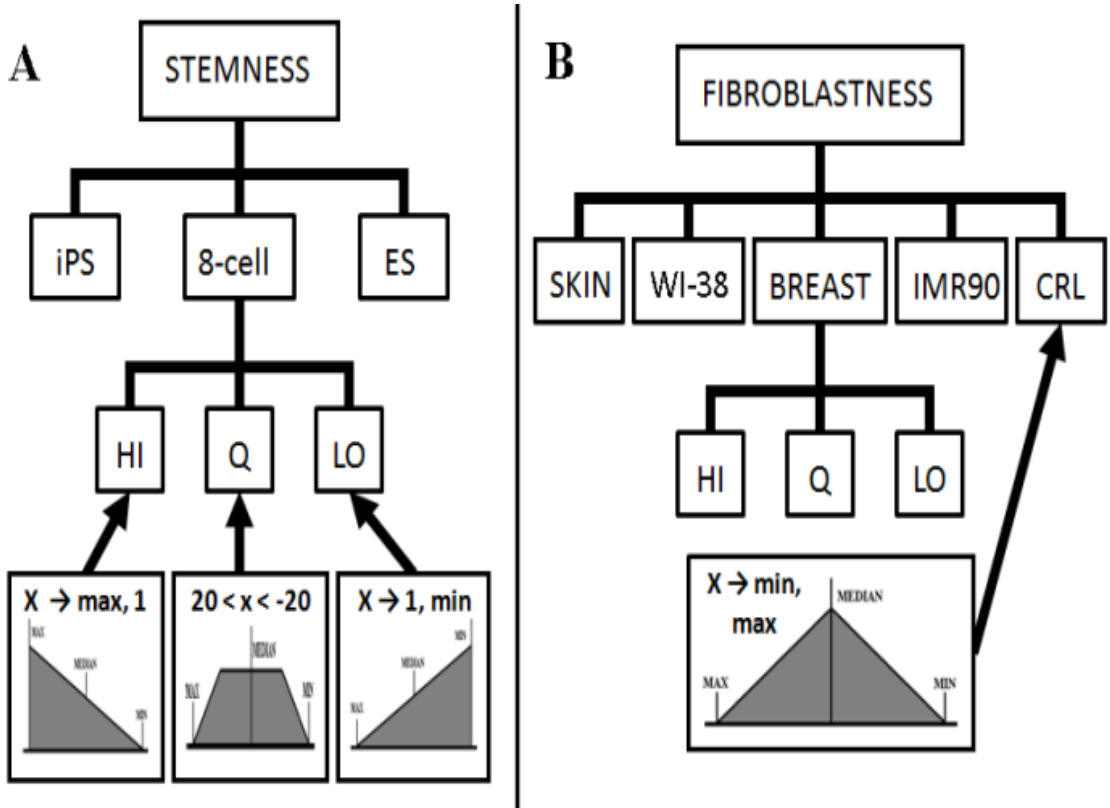

**Figure S2.** Schematic showing the fuzzy classification scheme for the fibroblast (A - left) and pluripotent (B – right) cell datasets.

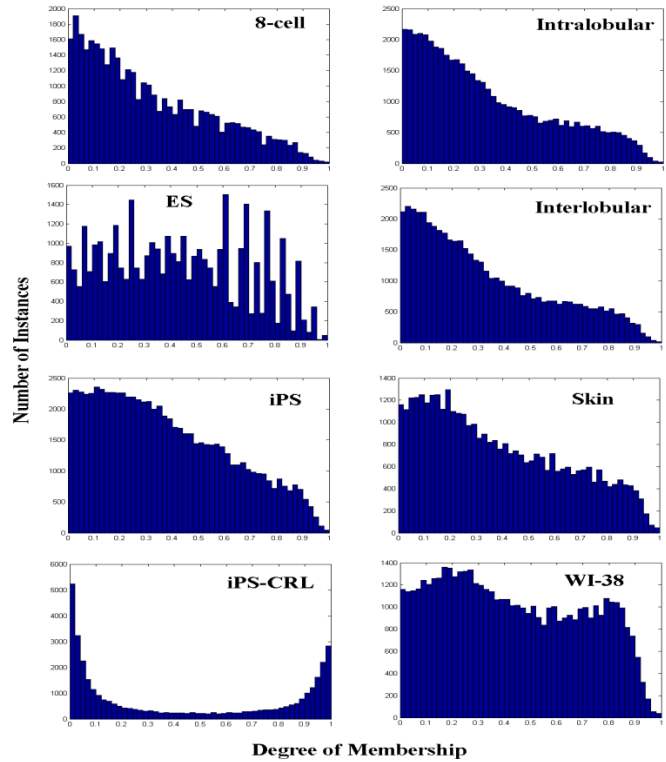

**Figure S3. Fuzzy (soft) classifier results for classification within cell lines for kernel using dual criterion. Histogram of 50 bins.**

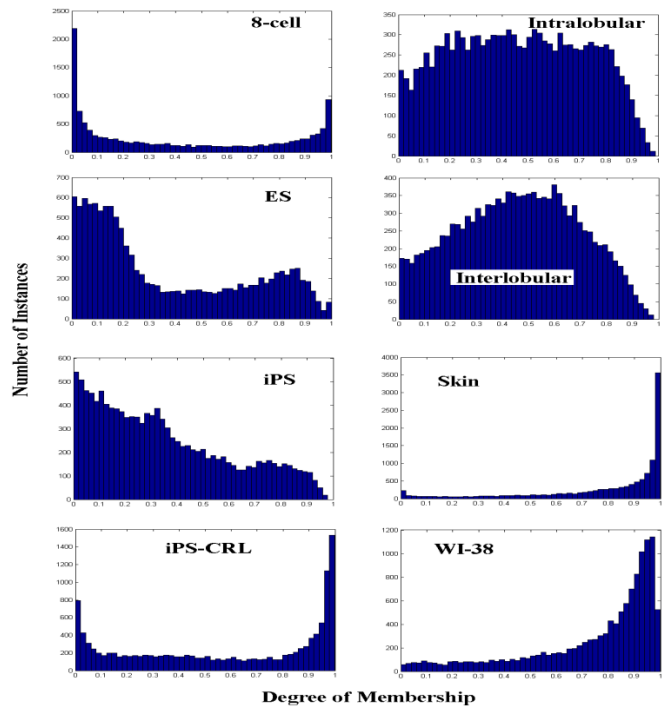

**Figure S4. Fuzzy (soft) classifier results for classification between cell lines but within a cellular state for kernel using dual criterion. Histogram of 50 bins.**

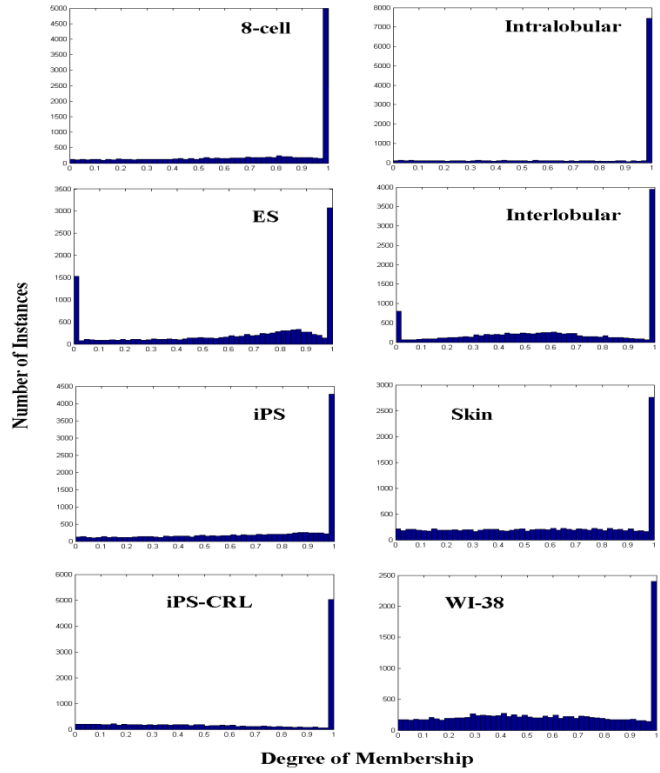

**Figure S5. Fuzzy (soft) classifier results for classification between cellular states for kernel using dual criterion. Histogram of 50 bins.**

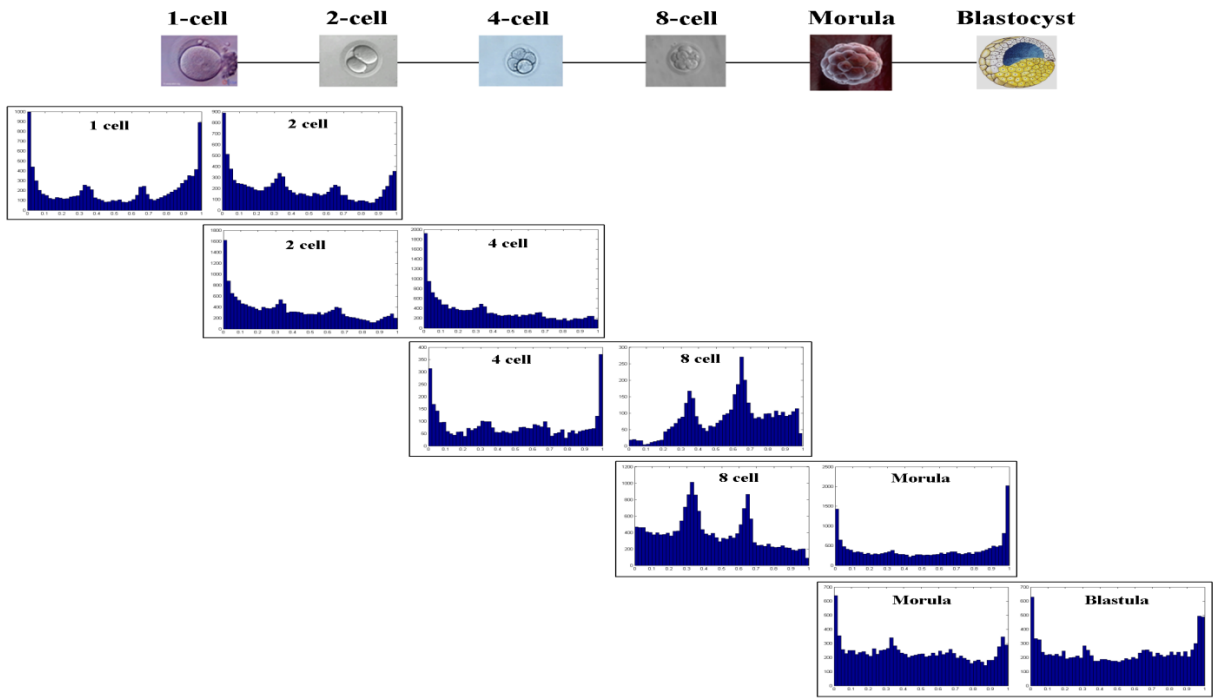

**Figure S6. Fuzzy (soft) classifier results for classification between cell lines (pairwise comparisons) within Embryo time-series for kernel using dual criterion. Histogram of 50 bins.**

18